# HEAT TRANSFERS IN A DOUBLE SKIN ROOF VENTILATED BY NATURAL CONVECTION IN SUMMER TIME

P. H. Biwole[a, b*], M. Woloszyn[a], C. Pompeo[b]

[a] Centre de Thermique de Lyon, CNRS-UMR 5008, Université de Lyon, INSA-Lyon, F-69621 Université Lyon I,F-69622, France.
[b] Centre Scientifique et Technique du Bâtiment, 24 rue J. Fourier, 38400 St-Martin d'Hères.

**ABSTRACT.**

The double-skin roofs investigated in this paper are formed by adding a metallic screen on an existing sheet metal roof. The system enhances passive cooling of dwellings and can help diminishing power costs for air conditioning in summer or in tropical and arid countries. In this work, radiation, convection and conduction heat transfers are investigated. Depending on its surface properties, the screen reflects a large amount of oncoming solar radiation. Natural convection in the channel underneath drives off the residual heat. The bi-dimensional numerical simulation of the heat transfers through the double skin reveals the most important parameters for the system's efficiency. They are, by order of importance, the sheet metal surface emissivity, the screen internal and external surface emissivity, the insulation thickness and the inclination angle for a channel width over 6cm. The influence of those parameters on Rayleigh and Nusselt numbers is also investigated. Temperature and air velocity profiles on several channel cross sections are plotted and discussed.

**Keywords:** Double-skin roof; Radiation; Natural convection; Passive cooling; Solar loads.

[*] Corresponding author: Centre de Thermique de Lyon, CNRS-UMR 5008, INSA de Lyon, bât. Sadi Carnot, 9 rue de la Physique, 69621 Villeurbanne cedex, France.
Tel: +33 472438468; Fax: +33 472438522.
Email address: pascal-henry.biwole@insa-lyon.fr

**Nomenclature**

| | |
|---|---|
| a | Thermal diffusivity (m²/s) |
| Cp | Specific heat capacity (J/kg.K) |
| e | Roof width (m) |
| E | Solar radiation (W/m²) |
| F | Buoyancy force (N) |
| g | Gravitational acceleration (m/s²) |
| H | Roof length (m) |
| ha | Convective heat transfer coefficient in the channel (W/m².K) |
| he | Convective heat transfer coefficient between outdoor air and screen (W/m².K) |
| hi | Convective heat transfer coefficient between indoor air and ceiling (W/m².K) |
| n | Vector normal to a surface oriented outwardly. |
| Nu | Nusselt Number |
| P | Pressure (Pa) |
| Ra | Rayleigh Number |
| Re | Reynolds Number |
| T | Temperature (K) |
| Te | Outdoor air temperature (K) |
| Ti | Indoor air temperature (K) |
| $T_{sky}$ | Sky temperature (K) |
| u | Air velocity field (m/s) |

Greek symbols:

| | |
|---|---|
| α | Absorption coefficient for solar radiation (-) |
| β | Coefficient of thermal expansion (1/K) |
| ε | Surface emissivity (-) |
| λ | Thermal conductivity (W/m.K) |
| μ | Dynamic viscosity (Pa.s) |
| ν | Cinematic viscosity (m²/s) |
| Φ | Heat flux density (W/m²) |
| ρ | Density (kg/m³) |
| σ | Stefan-Boltzmann constant: σ =5.67e-8 W/(m²K⁴) |
| θ | Roof slope (°) |

Subscripts

| | |
|---|---|
| se | Characteristic at the screen external surface |
| si | Characteristic at the screen internal surface |
| sm | Characteristic at the sheet metal surface |

# 1. INTRODUCTION.

According to the French Environment and Energy Management Agency [1] (ADEME), the air conditioning of dwellings in summer increases the annual power consumption by more than 500KWh per habitation while the invoice rises by 10 percent. Highly reflective and ventilated double-skin roofs can be used to improve passive cooling in summer. In such a system radiation heat transfer is highly coupled with natural convection, and there is a need for a better understanding of the heat losses through double-skin roofs. Thanks to its high reflectivity, the screen must repel a large amount of solar radiation. The heat absorbed and reemitted toward the habitation is warded off by the buoyancy forces between the screen and the sheet metal: This is ventilation by natural convection. The remaining conductive heat from the sheet metal is hampered by the insulation layer underneath.

In this work, heat and mass transfers in the tilted channel were numerically simulated and experimentally validated. Averaged Nusselt and Rayleigh numbers in the cavity, airflow rate, air temperature, air velocity in the cavity, and total heat flux through the ceiling were investigated. The aim was to determine the paramount factors for the double-skin roof efficiency, in terms of protecting the dwelling from solar loads.

After an overview of the literature on the subject, the model of the double-skin roof is described. Then, the equations governing heat and mass transfers through the roof and the simulated boundary conditions are presented. For validation purposes, results from the numerical model are compared to experimental measurements. Once validated, the numerical model is used to carry out a parametric study of a full-scale double roof structure.

# 2. LITERATURE REVIEW

In 1997, Fracastoro [2] studied the idea of reducing heat gain in dwellings by using under-roof cavities. He presented a numerical model for steady-state thermal analysis of ventilated and unventilated light roofs. In order to predict the performances of light roofs, the output data were air temperature distributions, surface temperatures, heat fluxes and air flow rates. Predictions were made with good accuracy but the algorithm had to be initialized with near-of-the-solution values.

In 1998, Lacena-Neildez [3] made a numerical and experimental simulation of heat transfers in innovative building components. She focused on the design of double-skin metal roofs for warm countries. She studied the influence of roof length and width, slope and external wind velocity on the passive cooling efficiency. She devoted greater attention to the influence of surface emissivities which proved to be the major parameter for efficiency. She tried different innovative paintings on the double-skin roof external surface. But her model was one-dimensional. Poquette [4] and Bolsée [5] took over in 2002 and 2003 and tried to build a 2-D numerical model but failed to reach steady and reliable convergence. In 2002, the French Scientific and Technical Centre for Building Research conducted a series of measurements on an experimental double-skin roof [6]. In 2003, Miranville et al. designed a multi-zone model of a double skin roof incorporating a radiant barrier system. After adjusting some heat transfer coefficients, comparison with experimental measurements was made with good agreement, thus validating the model [7]. In 2007, Chang, Chiang and Lai [8] evaluated experimentally the energy savings achieved by incorporating a radiant barrier system in a double-skin roof.

Regarding heat transfer coefficients within tilted cavities, Elenbaas [9] in 1942 built relations for natural convection between isothermal parallel plates at the same temperature. He also defined an optimal width for buoyancy driven thermal transfers. In 1985, Azevedo and Sparrow [10] experimentally built relations for the Nusselt and the Rayleigh number in differentially heated open-ended and tilted parallel plates. They only dealt with angles varying from 45 to 90° (vertical). One can notice a sensible difference when comparing their results to those obtained by Elenbaas, particularly regarding the influence of the channel's width on the correlations.

Concerning the external heat transfer coefficients, in 1954, Mac Adams [11] established a relation between the external wind velocity and the convective heat transfer coefficient. In 1986, Chen found some correlations for tilted isothermal (or with a constant surface flux) lonely plates but the correlations did not include wind's velocity.

In 1998, Sandberg [12] worked on solar chimneys and Aboulnaga [13] on the impact of solar cells on double-skin cavities. Though their issues are similar to ours, these studies do not directly deal with open-ended double-skin roofs.

## 3. NUMERICAL SIMULATION DESCRIPTION

### 3.1. Case description

The geometry of the model is presented in figure 1. The physical characteristics of the simulated materials are presented in table 1.

### 3.2. Governing equations.

#### 3.2.1. Within solid parts:

In solid parts (metal, insulation and plaster), heat transfers are dominated by conduction. The following form of the heat diffusion equation can be considered:

$$\rho C_p \frac{DT}{Dt} + \nabla.(-\lambda \nabla T) = Q + \mu\Phi \qquad (1)$$

where $\frac{D}{Dt}$ is the Lagrangian derivative, $Q$ the heat source term, and $\mu\Phi$ the heat gained from viscous frictions. Within solids, $Q = \Phi = 0$ then

$$\rho C_p \frac{\partial T}{\partial t} = \lambda \Delta T \qquad (2)$$

#### 3.2.2. On surfaces:

Above the screen, we take into account solar radiation and convection terms:

$$-\lambda_{screen} \frac{\partial T}{\partial y} = h_e(T_e - T_{se}) + \varepsilon_{se}\sigma(T_{sky}^4 - T_{se}^4) + \alpha_{screen}E(t) \qquad (3)$$

where $T_{se}$ is the temperature of the screen's external surface. The screen's external surface emissivity $\varepsilon_{se}$ is defined assuming that sky is a black body viewed under a view factor of 1. We did not measure $\varepsilon_{se}$, $\lambda_{screen}$ and $\alpha_{screen}$. They were given by the manufacturer as physical properties of the sheet metal.

Beneath the screen, we take into account convection with channel's air and long wave radiation with sheet metal:

$$-\lambda_{screen} \frac{\partial T}{\partial y} = h_a(T_{channel} - T_{si}) + r(T_{sm}^4 - T_{si}^4) \qquad (4)$$

where $T_{sm}$ is the temperature of the sheet metal cavity-side surface and $T_{si}$ the temperature of the screen's internal surface. As given by [16] and fully developed in section 3.3., the radiation coefficient r is a function of the surface emissivities of the opposing surfaces and of σ, assuming a view factor of 1 between parts of the screen and opposite parts of the sheet iron.

We have the same form above sheet metal:

$$-\lambda_{sheet}\frac{\partial T}{\partial y} = h_a(T_{channel} - T_{sm}) + r(T_{si}^4 - T_{sm}^4) \qquad (5)$$

Beneath the ceiling, we neglect long wave radiation with room. Such an approximation is done because of the feeble impact of the long wave radiation on the air layer's temperature in the double skin roof. Therefore, we can write:

$$-\lambda_{plaster}\frac{\partial T}{\partial y} = h_i(T_i - T_{ci}) \qquad (6)$$

where $T_{ci}$ is the ceiling's indoor surface temperature.

### 3.2.3. In the channel:

Heat equation (1) is coupled with Navier-Stokes equations for incompressible fluids. The velocity field $\vec{u}$ in the Lagrangian derivative is given by the continuity equation and the momentum conservation equation. Density variations of the air in the channel are supposed negligible when compared to the variation of air velocity. Therefore, the continuity equation reads:

$$\nabla \cdot \vec{u} = 0 \qquad (7)$$

The momentum conservation equation reads, in the stationary case:

$$\rho \vec{u} \cdot \nabla \vec{u} = \nabla \cdot (\mu(\nabla u + (\nabla u)^T) - \nabla P + F \qquad (8)$$

Given the Boussinesq approximation:

$$\rho = \rho_e(1 - \beta(T - T_e)) \qquad (9)$$

The buoyancy force $\vec{F}$ reads:

$$\vec{F} = -\rho_e(1 - \beta(T - T_e))\vec{g} \qquad (10)$$

### 3.3. Choice of heat transfer coefficients.

As shown in the Literature review chapter, many studies have been conducted on heat transfer coefficients for heated plates and channels. But most correlations found involve plates' temperature which, in our case, is not known beforehand but is an output of the computation. Consequently, upon the roof, we used the following convective heat transfer coefficient $h_e$ as given by Mac Adams [11]:

$$h_e = 5.7 + 3.8V \quad W/(m^2.K) \tag{11}$$

In the channel, $h_a$ was calculated as the average value between the relations given by the French thermal regulation [14] for a vertical cavity and a horizontal cavity:

$$h_a = \frac{1}{2}\left[\max\left(0.12 \cdot d^{-0.44}; \frac{0.025}{d}\right) + \max\left(1.25; \frac{0.025}{d}\right)\right] \quad W/(m^2.K) \tag{12}$$

where d is the hydraulic diameter. Here, d is the width of the air cavity.

hi in eq. (6) is the convective heat transfer coefficient between indoor air and ceiling. As previously said, long wave radiation with the room is neglected due to its feeble impact on the roof temperature. hi is also calculated as the average value between the relations given by the French thermal regulation [14] for an horizontal heat flux ($h = 2.5 W/m^2.K^{-1}$) and for a downward heat flux ($h = 0.7 W/(m^2.K^{-1})$):

$$h_i = 1.6 \quad W/(m^2.K)$$

Assuming a view factor of 1 between parts of the screen and opposite parts of the sheet iron, the constant r for radiation heat transfers in the channel is given by Incorpera and Dewitt [16] and can be demonstrated as follows:

Let be $R_{sm} = \sigma.\varepsilon_{sm}(T_{sm})^4$ the initial radiation flux going downward from screen to sheet iron and $R_{si} = \sigma.\varepsilon_{si}(T_{si})^4$ the initial radiation flux going upward from sheet iron to screen.

$R_{sm}$ and $R_{si}$ are subjected to many reflexions. Let be $\rho_{sm}$ et $\rho_{si}$ the reflectivities of the cavity-side face of the screen and of the cavity-side face of the sheet iron, respectively. The following chain reaction occurs: the screen emits a flux $R_{sm}$ which provokes a reflective flux $\rho_{si} R_{sm}$ at the sheet iron surface, then another reflexion $\rho_{ie} \rho_t R_e$ on the screen and so on. All the same, the sheet iron emits $R_{si}$, which is reflected into $\rho_{sm} R_{si}$ and go upward again as $\rho_{si} \rho_{sm} R_{si}$…

All in all, if we assume a view factor of 1 between parts of the screen and opposite parts of the sheet iron, the descending and ascending radiation heat flux can be written as:

$$\Phi\downarrow = R_{sm} + \rho_{si}\rho_{sm}R_{sm} + \rho^2_{si}\rho^2_{sm}R_{sm} + \ldots + \rho_{sm}R_{si} + \rho^2_{sm}\rho_{si}R_{si} + \ldots \quad (13)$$

$$\Phi\uparrow = R_{si} + \rho_{si}\rho_{sm}R_{si} + \rho^2_{si}\rho^2_{sm}R_{si} + \ldots + \rho_{si}R_{sm} + \rho_{sm}\rho^2_{si}R_{sm} + \ldots \quad (14)$$

The net heat flux of a surface i is defined as the heat flux emitted or leaving the surface minus the heat flux absorbed or arriving by the same surface. In our case, the net descending flux is obtained by doing (13)-(14). One obtains:

$$\Phi\downarrow net = (1+\rho_{si}\rho_{sm} + \rho^2_{si}\rho^2_{sm}+\ldots)(R_{sm}(1-\rho_{si})+(R_{si}(\rho_{sm}-1))) \quad (15)$$

This is a geometric progression with $\rho_{si}\rho_{sm}$ as the common ratio number. However, the absorptivity, the reflectivity and the transmitivity of a material subjected to radiation are linked by the relation:

$$\alpha + \tau + \rho = 1 \quad (16)$$

All the surfaces being opaque, $\tau = 0$ and Kirchoff's law for grey bodies gives $\varepsilon = \alpha$. Therefore, eq. (16) becomes:

$$\varepsilon + \rho = 1.$$

Then

$$\Phi\downarrow net = [1/(1- \rho_{si}\, \rho_{sm})].\varepsilon_{si}\, \varepsilon_{sm}\sigma\, (T_{sm}^4 - T_{si}^4)$$
$$= [\varepsilon_{si}\varepsilon_{sm}/(\varepsilon_{sm} + \varepsilon_{si} - \varepsilon_{si}\varepsilon_{sm})]\, \sigma\, (T_{sm}^4 - T_{si}^4)$$

$$\Phi\downarrow net = r\, (T_{sm}^4 - T_{si}^4) \quad (17)$$

where

$$r = \left[\frac{\varepsilon_{si}\varepsilon_{sm}}{\varepsilon_{si} + \varepsilon_{sm} - \varepsilon_{si}\varepsilon_{sm}}\right]\sigma \quad W/(m^2K^4) \quad (18)$$

### 3.4. Boundary conditions.

Sky temperature used in equation (3) is defined as the temperature of a black hemisphere absorbing the same radiation flux as the sky. We kept the correlations given by Swinbank [15]:

$$T_{sky} = 0.0552\, (Te)^{1.5}\ (K) \quad \text{for a clear sky.} \quad (19)$$
$$T_{sky} = Te\ (K) \quad \text{for a cloudy sky.}$$

The thermal boundary conditions in the channel are as follows: External temperature Te imposed at the entry of the channel. To achieve numerical convergence with an unknown exit temperature, a condition on the convective flux was specified at the channel exit, meaning that in (1),

$$-k\nabla T \cdot n = 0 \qquad (20)$$

The total heat flux is then limited to its convective part:

$$q \cdot n = \rho C_p T u \cdot n \qquad (21)$$

A no-slip condition of the fluid on the side plates of the channel was imposed as the dynamic boundary condition. Therefore, the air velocity equals zero on the plates. In order to numerically simulate the exit gap, the total force on the exit boundary was set to zero. Therefore in equation (5),

$$T = (-pI + \eta(\nabla u + (\nabla u)^T))n = 0 \qquad (22)$$

Thus, inertia forces are balanced by buoyancy forces.

In order to help numerical convergence and simulate efficiently adverse summer condition with no wind, the air velocity at the cavity entry was initialised as nil. The screen, cavity and sheet iron initial temperature is Te. The insulation and plaster initial temperature is Ti.

**3.5 Grid and computation.**

The computation was done with a finite elements method. For a better computation of near-the-wall heat and mass transfers, the grid was quadrilateral and refined near the borders of the channel. The analysis was stationary. A refining of the mesh showed an augmentation of the temperature output between 0 and 0.4 K and a 0.012 m/s increase of the channel's air velocity. Nevertheless, the results did not vary anymore beyond a 100 000-degree of freedom mesh. The grids used possessed around 148650 degrees of freedom and 12710 finite elements each.

In order to respect the full coupling between heat and mass transfers' non linear equations, the equations presented above (in channel and solids) were solved simultaneously to produce temperature and velocity fields.

## 4. ELEMENTS OF VALIDATION

### 4.1. Experimental setup.

Two experimental set up were erected at the French Scientific and Technical Centre for Building Research (CSTB). The results were confronted to numerical simulation. The first experiment consisted of a 1m large and long, 1.2mm-wide sheet iron laid over a 3cm-wide insulation with an air gap of 15mm between sheet and insulation. In the second experiment, an additional sheet metal was placed over the first one. In order to have the same aspect ratio as a 4m-long double-skin roof with a 10cm cavity, both sheets were separated by 2.5cm-wide wood rafters (see figure 2). The mere presence of the wood rafters helped the chimney effect by accentuating the pressure gradient between the air in the channel and the air outside.

As sketched out in figure 2, T-type thermocouples, named T1 to T4 were placed in the middle and both faces of each sheet metal used. T4 was placed on the external face of the upper sheet, and T3 on the cavity-side face of the same sheet. T2 was placed on the cavity side face of the lower sheet whereas T1 was facing insulation. Given the high thermal conductivity of the metal forming the parts of the roof and the stationary regime of simulation, no thermocouple was placed on the edge of those parts. The T-type thermocouples used were made of copper. They had a precision of ± 0.2°C and a temperature range of -200°C to 300°C. Some shots of the experimental set-up are shown in figure 3.

The measurements were performed every hour from 9 am to 16 pm. Air velocity, ambient temperature and solar radiation were measured every ten minutes by a weather station installed in the immediate vicinity of the experimental set up. The results of measurements taken on one typical sunny day are shown in figures 4 and 5 respectively for the simple roof measured data and for the double-skin roof measured data. Physical characteristics of materials in use are the same as in Table 1. In figure 5, we notice a sensible decrease of the temperature on the roof when we create the double skin structure by adding the second sheet iron.

### 4.2. Comparison with numerical simulation.

Dynamic comparison between predictions and measurements are performed on an hourly basis using the weather conditions registered on the 2$^{nd}$ of August 2006 inside the Scientific and Technical Centre for Building Research, Grenoble, France.

Due to its easy access, the validation criteria used is temperature of screen and sheet iron taken on the four thermocouples T1 to T4. T1 being very close to T2 (respectively T3 to T4), only T2 and T4 are represented in figure 6. The figure shows a noticeable difference between numerical and experimental results at the early and late hours, and likeness of results for the middle part of the day. This is due to the thermal inertia of the sheet metal which is not taken into account by numerical simulation since the regime simulated is stationary. Heat variations of the sheet metal are less rapid in the real case. Once the sheet metal heated, numerical and experimental results witness the same variations with pretty small differences of temperatures as shown in table 2.

Since we placed ourselves in the most unfavourable case of a hot and sunny afternoon with no wind, the simulations were performed in stationary regime. This interval of validity (from 11 am to 3 pm as shown in figure 6) for our validation criteria and this level of accuracy are sufficient to discuss the relative importance of the double-skin-roof parameters on its efficiency.

## 5. RESULTS

### 5.1. List of numerical simulations

Once validated, the numerical model was used to study a full scale double-skin roof. The studied roof was 4m long and 1m large. An extensive parametric study was performed by making vary the roof slope, the cavity width, the insulation thickness, the solar radiation, and the surfaces' emissivities. The different cases simulated are described in Table 3 and the results are presented and discussed in the following sections.

The constants used for boundary conditions and air physical characteristics are presented in Table 4:

## 5.2. Results and discussion.

### 5.2.1. Temperature and air velocity fields

Temperature and air velocity profiles on several cross sections of the channel where plotted under the conditions of simulation case (q) (see fig. 7). The more and more V-shaped profile of the temperature along the channel is due to the warming process of the air through the cavity. We notice an air layer in the middle of the channel which remains at the initial temperature. It wouldn't have been the case if the cavity was thinner or if the roof was longer because the thermal boundary layers of the two plates would have made contact. We also notice that screen and sheet metal's surface temperature remain nearly constant over the whole channel length. Screen's temperature varies from 338.0K to 340.5K whereas sheet metal's temperature varies from 319.9 to 323.6K, the lower values being observed near the channel's entry. The difference is rather small regarding the size of the roof (4m). This fact justifies the use of a comprehensive parameter such as the total heat flux on the ceiling to measure the system efficiency.

The air velocity profile on every cross section is parabolic, with a zero value on both plates due to the no-slip boundary condition. Maximum speed is observed near the screen. This is caused by the buoyancy force which is proportional to the difference between plate's temperature and outdoor temperature. $\vec{F}$ is largest near the screen, for screen is the hottest of the two plates. In spite of its warming, the air gains speed throughout the channel. We noticed maximum speed of 0.89m/s at the entry and 0.94m/s at the exit of the channel.

Similar profiles were obtained for every simulation. Therefore in the following, no more details are given and only averaged values, representative of each case are presented.

*5.2.2. Impact of inclination angle*

The average Nusselt number describing transfers between channel's air and screen was computed using equations (23) and (24)

$$\overline{N_u} = \frac{e}{\lambda}\overline{h} \qquad (23)$$

and

$$\overline{h} = \frac{1}{H}\int_0^H \frac{q_x}{(T_x - 297)}dx \qquad (24)$$

$q_x$ being the convective heat flux exchanged between the internal screen surface and channel's air at x abscissa along the channel. The results for case (g) are presented in figure 8. The averaged Nusselt number increases when the roof inclination angle increases.

Natural convection is made easier by a higher slope. This result is coherent with equation (10) which shows that after projection on x-axis, the buoyancy force varies proportionally with $\sin(\theta)$. Simulation (g) was used to investigate the impact of the angle of inclination on the mean temperature in the cavity and the mean air velocity at the channel exit (see figure 9). We found that cavity's temperature decreases and air velocity increases when the angle rises.

*5.2.3. Impact of channel width and insulation thickness*

Results of simulations (a) to (f) are plotted in figure 10, demonstrating that the average Nusselt (eq. 23 and 24) and Rayleigh (eq. 25) numbers in the channel increase when the channel's width increases. The Nusselt number is more cavity-width-sensitive whereas the Rayleigh number is more insulation-width-sensitive. For a fixed cavity width, the Rayleigh number increases when the insulation thickness diminishes. It can be explained by the fact that a diminution of insulation diminishes the sheet iron mean temperature by allowing heat to get through. Consequently, radiation heat transfer between screen and sheet metal rises.

Rayleigh number is computed using the following equation:

$$\overline{Ra} = \frac{g\beta(T_{si} - T_e)H^3}{\nu a} \qquad (25)$$

with $T_{si}$ being the mean temperature of the screen's internal surface.

Convective heat flux on the plaster ceiling was estimated using case (h). In this simulation the radiation part in equations (4) and (5) was removed. The value of convective heat flux reaching the ceiling was 0.284W for a 4cm-thick insulation, 7e-5W for a 6cm-thick insulation, and null for a 10cm-thick insulation. These results point out that convective heat flux from the screen is entirely evacuated outside the channel as soon as its width reaches 6cm.

Figure 11 shows that the sheet iron temperature increases by about 2°C for every 4cm of additional insulation layer. Given that the thermal resistance behind the sheet iron is increased, there is more heat absorbed. It was also found that for fixed cavity's width, mean and maximal screen's temperatures slightly decrease when insulation thickness increases. In the same time, ceiling temperature diminishes in a much pronounced way (6°C for every 4cm of additional insulation), as shown in figure 15. In figure 16 one can also notice a rise of mass flow in the channel. The averaged Nusselt number's dependence on the insulation width is not linear.

Figure 11 also shows that for fixed insulation thickness, mean and maximal screen and sheet metal temperatures slightly rise. This leads to an increase of the buoyancy force in the channel and consequently to a growth of mass inflow and outflow, as shows figure 12. What is more noticeable is that the total heat loss through the ceiling obtained by increasing insulation thickness is 10W per cm of insulation layer whereas the heat loss obtained by increasing the channel's width is only 0.8W per cm (see figure 13). Therefore one would rather increase the insulation thickness than increase the cavity width.

*5.2.4. Influence of radiation heat transfer*

The influence of natural convection on radiation heat flux is very low. Calculations show that the temperature of the sheet metal is nearly constant, whatever the channel's width. This is due to the overriding radiation heat transfer between screen and sheet metal. Therefore, sheet iron reflectivity and insulation thickness and quality play a key role in preventing heat gain by the dwelling.

Indeed, simulations (i) to (p), plotted in figure 14, showed that an 80% reduction (from 0.8 to 0.15) of either screen internal surface emissivity or sheet metal surface emissivity permits to halve (from 54.2W to 27.97W) the total heat flux crossing the ceiling. The same reduction of emissivity applied to the screen external surface only makes the flux to vary from 54.2W to 53.86W. When all surface emissivities vary from 0.8 to 0.15, the total heat flux through the ceiling decreases from 54.2W to 22.3W (see figure 14). Therefore, the most important surface emissivities to reduce are first the internal screen surface emissivity or the sheet metal emissivity, and last the external screen surface emissivity. As shown in figure 15, ceiling and sheet metal mean temperatures tend to decrease along with surface emissivities whereas screen temperature remains erratic.

**6. CONCLUSION**

In this work, a numerical model of a double-skin roof and the associated governing equations and boundary conditions were established. An experiment was set-up and used to validate the model. Then, a parametric study was performed on the validated numerical model with the purpose of determining the paramount factors for the roof efficiency and studying temperature and air velocity fields in the channel.

The average Nusselt number characterising heat transfers between channel's air and screen increases along with roof slope and insulation thickness. The average Rayleigh number increases when the insulation thickness diminishes. Temperature and air velocity profiles on channel cross sections are parabolic. Temperature of screen and sheet metal remain nearly constant, due to their high thermal conductivity. Maximum air speed is observed near the screen where the buoyancy force is the largest.

Numerical simulation showed that the efficiency of the double skin-roof relies on the following parameters: First the internal screen and the sheet metal surfaces' emissivities and second the external screen surface emissivity, all threes having to be as low as possible. Third, the insulation thickness has to be as high as reasonably possible. Fourth, the double skin width must be over 6cm and under 10cm. We did not witness noticeable improvement of efficiency for systems with channel's width over 10cm. Ventilation by natural convection is not significantly improved by cavities over 10cm. The convective part of the heat transfer remains at its minimal value whereas the radiation part remains high and predominant. The last parameter by order of importance for the system efficiency is the angle of inclination. Near-optimum efficiency can be easily reached with a double-skin structure featuring a 0.15-emissive sheet iron, a 5cm-thick insulation, a 10cm-wide cavity and an inclination angle over 30° from horizontal.

In spite of the good results obtained, some enhancements can be performed on the numerical model. Firstly, a more precise correlation linking roof slope to convective heat transfer coefficients could be used. Secondly, to gain access to heat transfers coefficients, plates' temperatures have to be known and set beforehand whereas they are supposed to be a result of the model. Further research has to be done in this area. To this regard, an iterative approach using temperatures of previous models to define heat transfers coefficients could yield more accurate results.

**Figure 1**

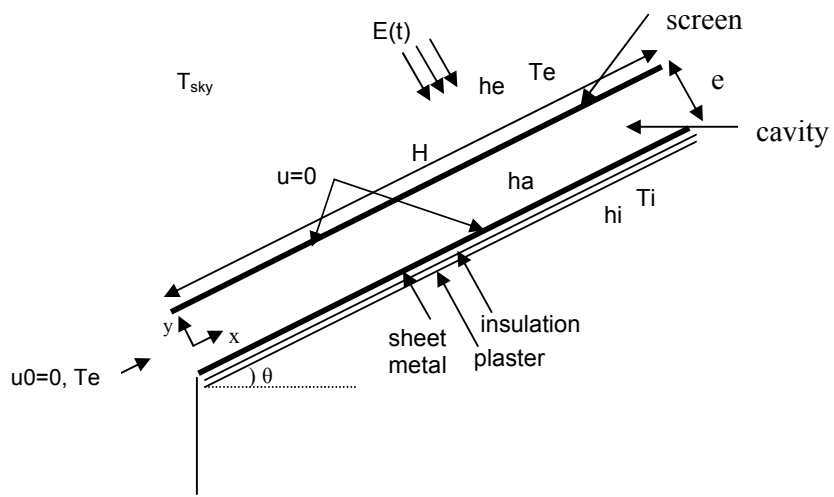

Fig. 1. Main elements of the numerical model

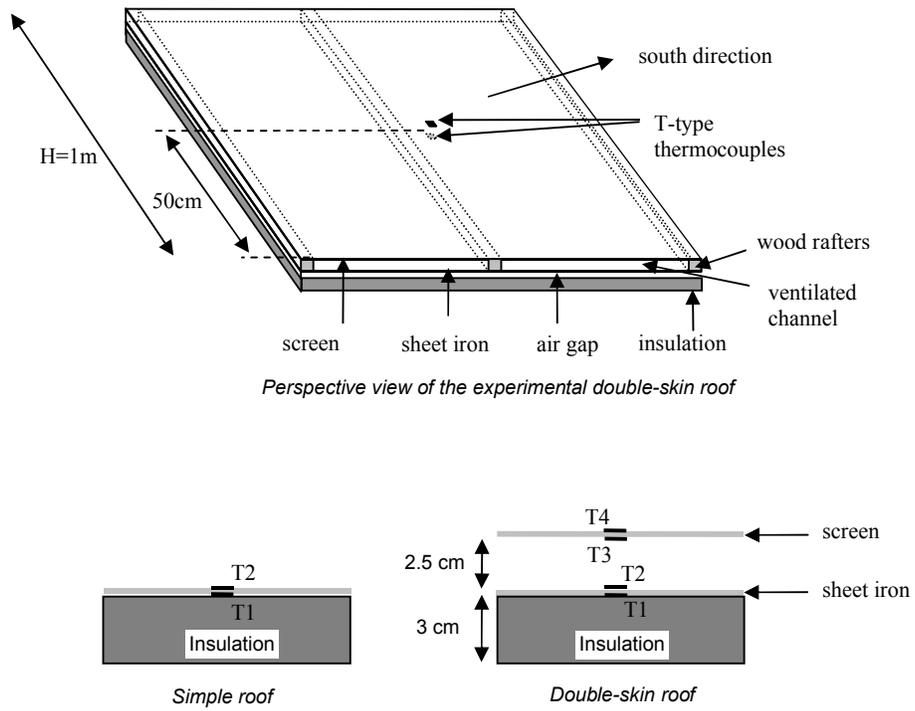

Fig. 2. Experimental setup.

**Figure 3**

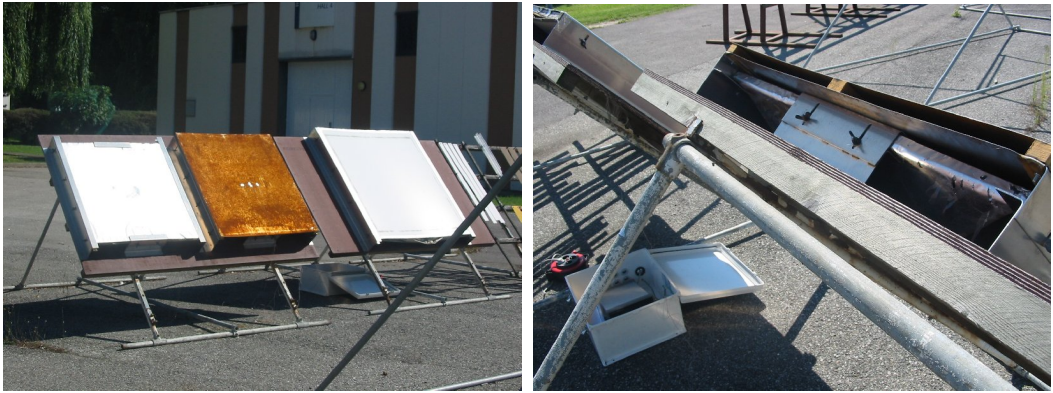

Fig 3. Shots of experimental setup

**Figure 4**

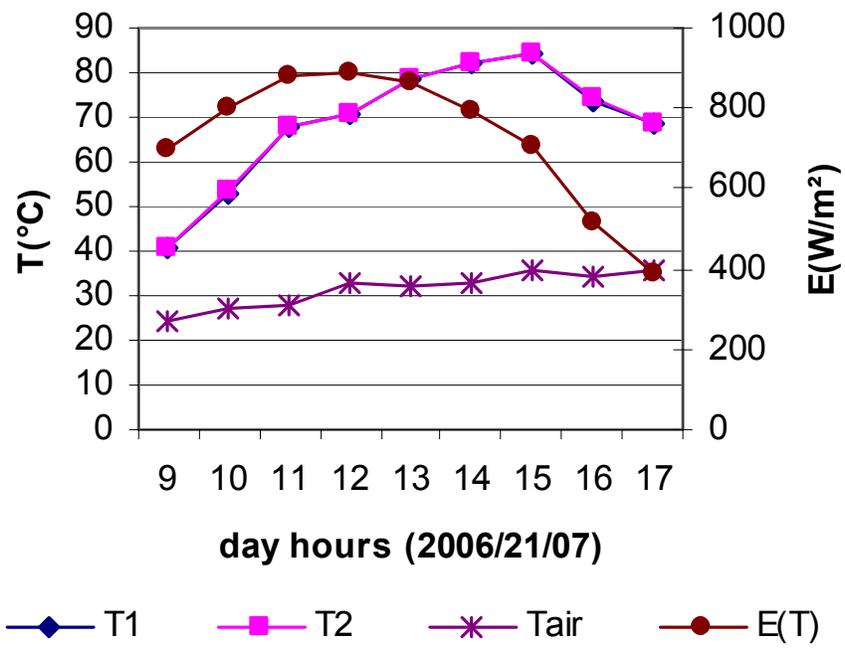

Fig. 4. Simple roof measured data

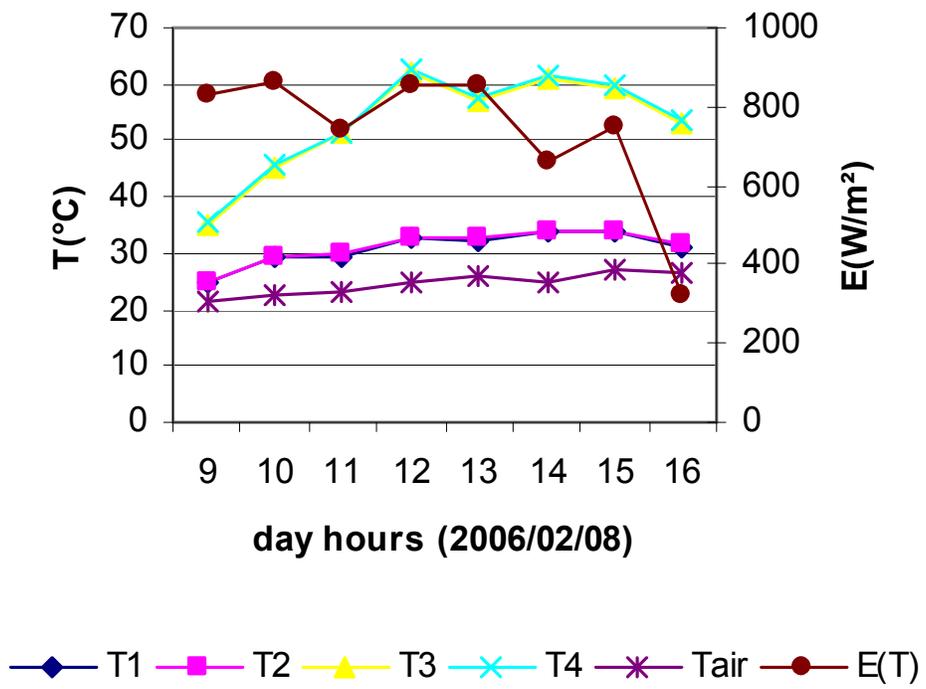

Fig. 5. Double-skin roof measured data

**Figure 6**

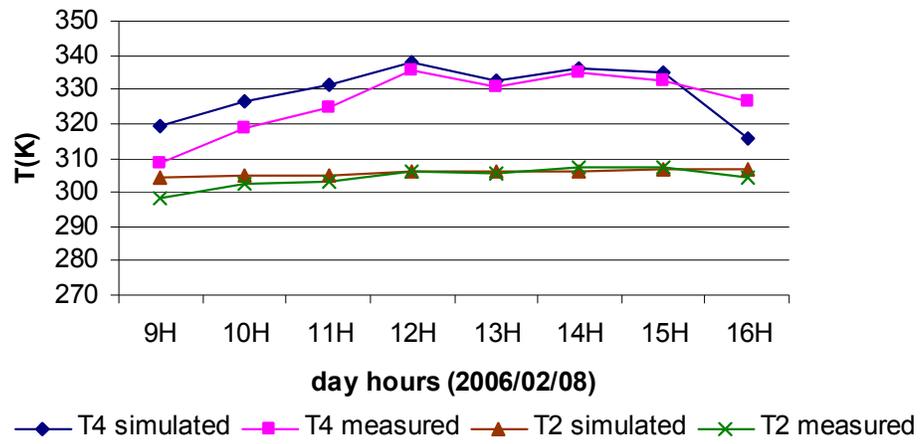

Fig. 6. Dynamic comparison between experimental and numerical results.

**Figure 7**

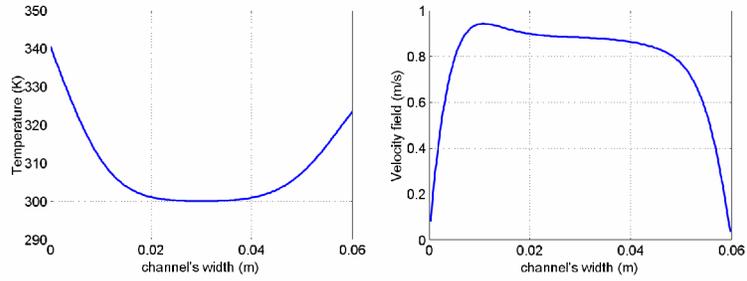

*Temperature and velocity profiles at 100% of the channel's length (exit)*

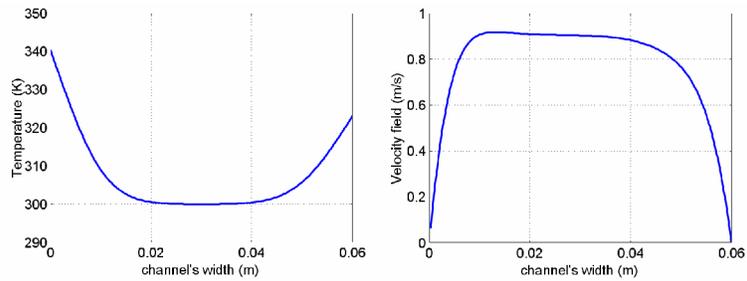

*Temperature and velocity profiles at 75% of the channel's length*

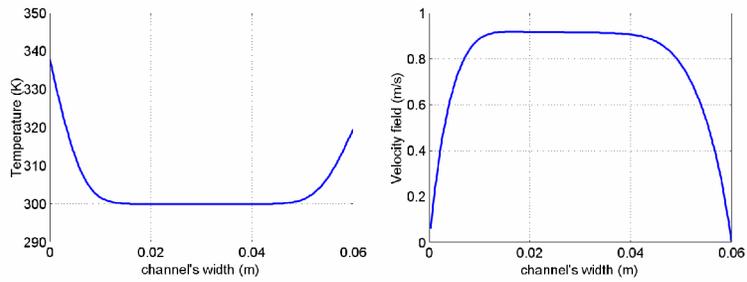

*Temperature and velocity profiles at 25% of the channel's length*

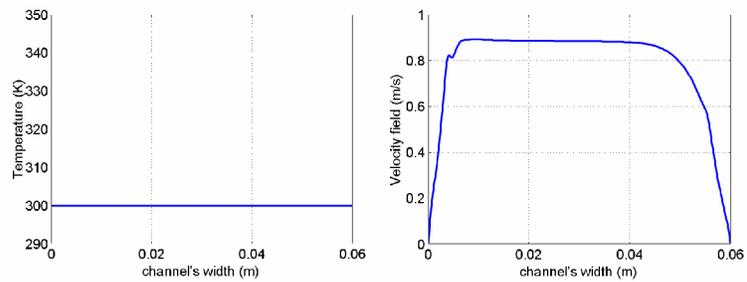

*Temperature and velocity profiles at 0% of the channel's length (entry)*

Fig. 7. Temperature and velocity's profiles on several channel cross sections.

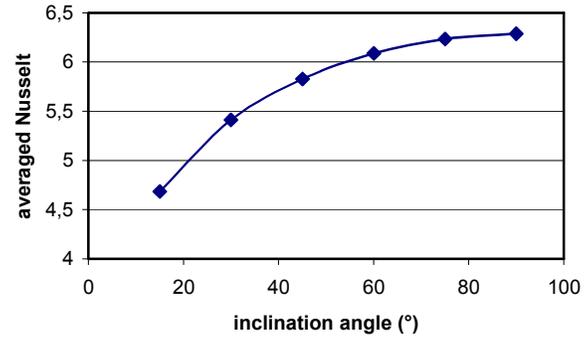

Fig. 8. Nusselt number induced by roof slope.

**Figure 9**

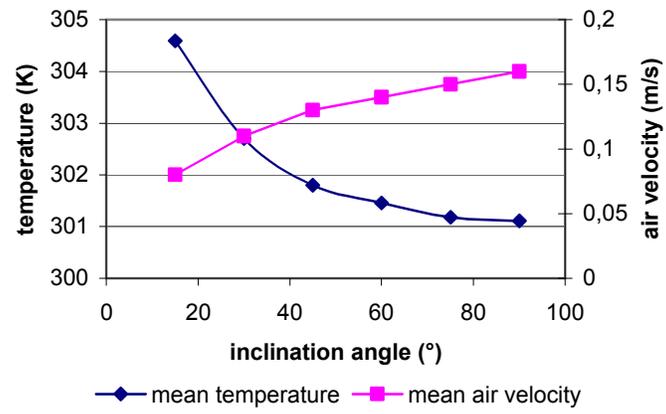

Fig. 9. Mean temperature and air velocity induced by inclination angle.

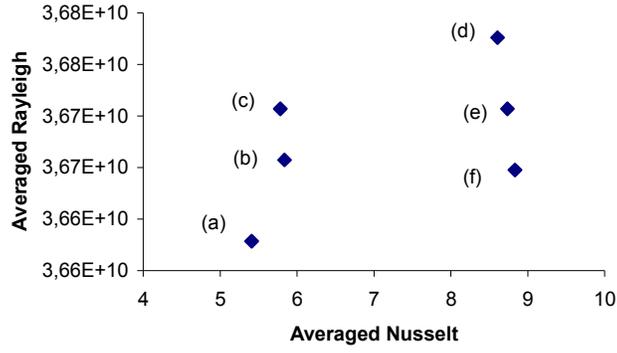

Fig. 10. Average Rayleigh number induced by average Nusselt number, cavity width and insulation thickness.

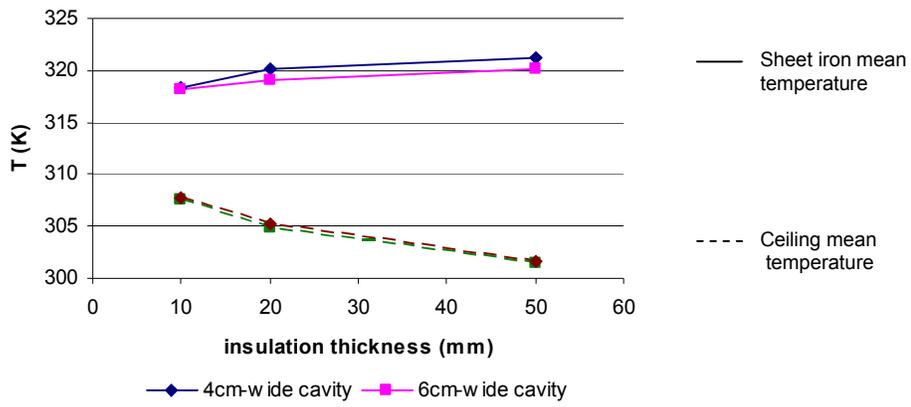

Fig. 11. Sheet iron and ceiling mean temperatures induced by insulation and cavity's width.

**Figure 12**

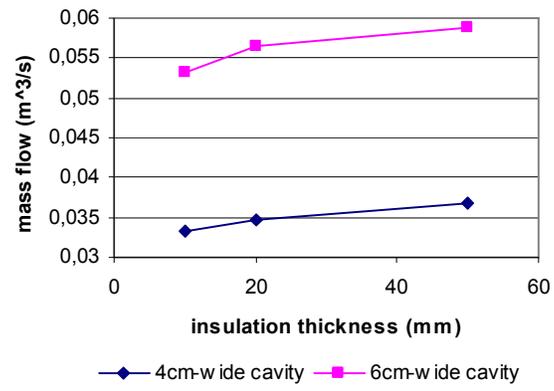

Fig. 12. Mass flow at the cavity exit induced by insulation and cavity's width

**Figure 13**

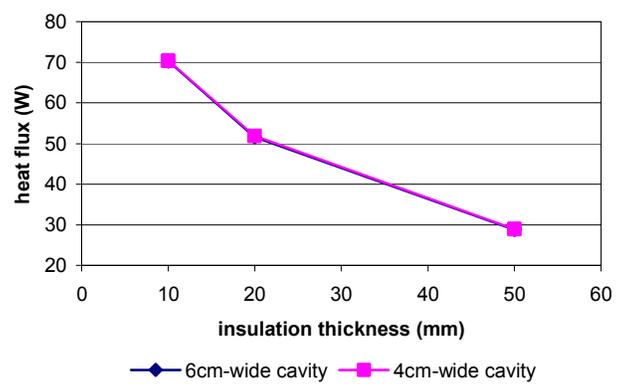

Fig. 13. Total heat loss through the ceiling induced by insulation and cavity's width

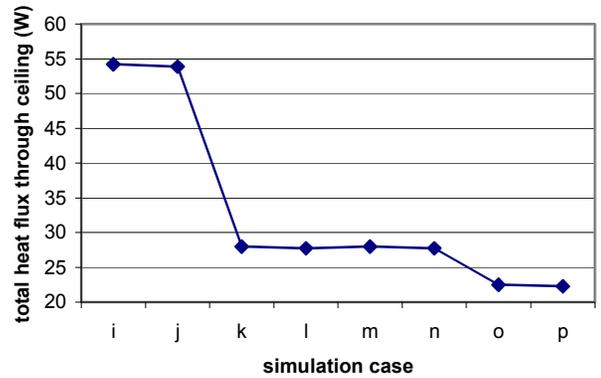

Fig. 14. Total heat flux through the ceiling induced by emissivities' variations.

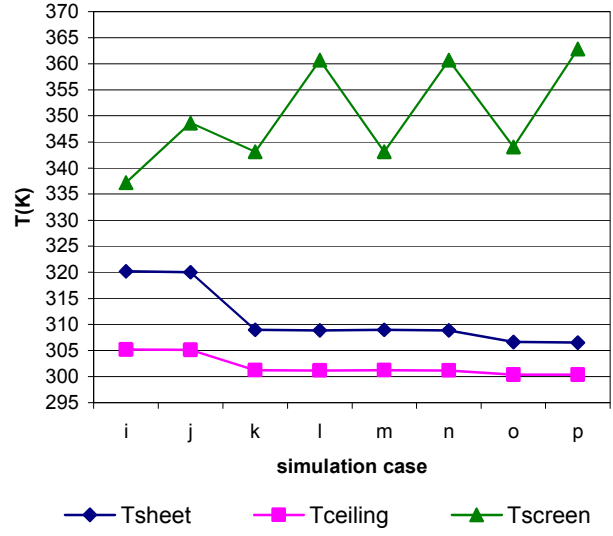

Fig. 15. Ceiling, sheet metal and screen temperatures induced by emissivities' variations.

**Table 1**

|  | Cp (J/kg.K) | λ (W/m.K) | ρ (Kg/m³) | Widths tested (m) |
|---|---|---|---|---|
| Sheet metal and screen | 880 | 160 | 2800 | 1e-3 |
| air cavity | 1008 | 0.025 | 1.23 | 0.1, 6e-2, 4e-2 |
| Insulation | 1030 | 0.04 | 15 | 1e-2, 2e-2, 5e-2 |
| Plaster | 1000 | 0.25 | 850 | 13e-3 |

Table 1. Materials physical characteristics.

**Table 2**

|  | Screen temperature | Sheet iron temperature |
|---|---|---|
| Experimental result | 330,95 K | 305,65 K |
| Numerical simulation result | 332,55 K | 306,25 K |

<u>Table 2.</u> Comparison between experimental and numerical results at 1 pm.

**Table 3**

| case | roof angle (°) | cavity width | insulation thickness | $\varepsilon_{se}$ | $\varepsilon_{si}$ | $\varepsilon_{sm}$ | Solar radiation (W/m²) |
|---|---|---|---|---|---|---|---|
| a | 30 | 4cm | 5cm | 0,8 | 0,8 | 0,8 | 800 |
| b | 30 | 4cm | 2cm | 0,8 | 0,8 | 0,8 | 800 |
| c | 30 | 4cm | 1cm | 0,8 | 0,8 | 0,8 | 800 |
| d | 30 | 6cm | 1cm | 0,8 | 0,8 | 0,8 | 800 |
| e | 30 | 6cm | 2cm | 0,8 | 0,8 | 0,8 | 800 |
| f | 30 | 6cm | 5cm | 0,8 | 0,8 | 0,8 | 800 |
| g | 15, 30, 45, 60, 75, 90 | 10cm | 2cm | 0,8 | 0,8 | 0,8 | 250 |
| h | 30 | 6cm | 2cm | 0,8 | 0,8 | 0,8 | 250 |
| i | 30 | 6cm | 2cm | 0,8 | 0,8 | 0,8 | 800 |
| j | 30 | 6cm | 2cm | 0,15 | 0,8 | 0,8 | 800 |
| k | 30 | 6cm | 2cm | 0,8 | 0,15 | 0,8 | 800 |
| l | 30 | 6cm | 2cm | 0,15 | 0,15 | 0,8 | 800 |
| m | 30 | 6cm | 2cm | 0,8 | 0,8 | 0,15 | 800 |
| n | 30 | 6cm | 2cm | 0,15 | 0,8 | 0,15 | 800 |
| o | 30 | 6cm | 2cm | 0,8 | 0,15 | 0,15 | 800 |
| p | 30 | 6cm | 2cm | 0,15 | 0,15 | 0,15 | 800 |
| q | 30 | 6cm | 5cm | 0,8 | 0,8 | 0,8 | 800 |

Table 3. List of cases numerically simulated.

**Table 4**

| Constant name | Te | Ts | Ti | he | ha | hi | g | $\rho_e$ | λ | B | μ | Cp |
|---|---|---|---|---|---|---|---|---|---|---|---|---|
| Constant value | 300 | 286,83 | 297 | 6 | 1,458 | 3,3 | 9,81 | 1,225 | 0,025 | 3,34E-3 | 1,62E-5 | 1008 |

Table 4. Constants used for numerical simulation.